\documentclass[twocolumn,floatfix,prl]{revtex4}%
\usepackage[dvipdfmx]{graphicx}%
\usepackage{amsmath}%
\usepackage{amsfonts}
\usepackage{amssymb}
\usepackage{mathrsfs}
\usepackage{color}
\usepackage{bm}

\def\s{{\sigma}}
\def\e{{\epsilon}}
\def\k{{ {\bm k} }}
\def\p{{ {\bm p} }}
\def\q{{ {\bm q} }}

\def\0{{ {\bm 0} }}
\def\w{{\omega}}
\def\a{{\alpha}}

\def\r{{ {\bm r} }}

\allowdisplaybreaks[4]

\begin{document}
\title{
Unconventional orbital-charge density wave mechanism
in transition metal dichalcogenide 1T-TaS$_2$
}
\author{
Toru Hirata, Youichi Yamakawa, Seiichiro Onari, and Hiroshi Kontani
}
\date{\today }

\begin{abstract}
The transition metal dichalcogenide 1T-TaS$_2$
attract growing attention because of the formation of 
rich density-wave (DW) and superconducting transitions.
However, the origin of the 
incommensurate DW state at the highest temperature ($\sim 550$K),
which is ``the parent state'' of the rich physical phenomena,
is still uncovered.
Here, we present a natural explanation for 
the triple-$\q$ incommensurate DW in 1T-TaS$_2$ 
based on the first-principles Hubbard model with on-site $U$.
We apply the paramagnon interference mechanism 
that gives the nematic order in Fe-based superconductors.
The derived order parameter has very unique characters: 
(i) the orbital-selective nature, and 
(ii) the unconventional sign-reversal in both momentum and energy spaces.
The present study will be useful for understanding 
rich physics in 1T-TaS$_2$, 1T-VSe$_2$, and other
transition metal dichalcogenides.

\end{abstract}

\address{
Department of Physics, Nagoya University,
Furo-cho, Nagoya 464-8602, Japan. 
}
 
\sloppy

\maketitle


The transition metal dichalcogenides (TMDs) 
provide a promising platform of the exotic low-dimensional
electronic states with strong electron correlation.
Among the TMDs, 
1T-Ta(S,Se)$_2$ exhibits very interesting electronic properties,
such as the metal-insulator transition
with the David-star formation as well as exotic superconductivity.
The electronic states are easily controlled by 
the gate-electric-field carrier doping 
\cite{gate-tunable1,gate-tunable2,gate-tunable3},
by changing the dimensionality
\cite{2D-state,monolayer,Dimer-Mott2020},
and by applying the pressure
\cite{Uwatoko,PT-phase}
and picosecond laser pulses
\cite{photoinduced}.

In 1T-TaS$_2$, at ambient pressure,
the incommensurate charge-density-wave (IC-CDW) 
appears as the highest transition temperature 
at $T=T_{\rm IC}\approx550$K \cite{Sipos}.
With decreasing $T$, the IC-CDW changes to the 
nearly commensurate (NC) CDW at $T_{\rm NC}\approx350$K,
and finally David-star commensurate (C) CDW 
appear at $T_{\rm C}\approx200$K successively.
This rich multistage CDW transition is suppressed under pressure,
and the superconductivity emerges at $T_{\rm SC}\lesssim10$K.
These exotic ordered states emerge under the IC-CDW state.
That is, the IC-CDW is the parent electronic state
of rich physics in 1T-Ta(S,Se)$_2$
\cite{K-Sun,R-Ang,M-Klanjsek,L-Ma,Y-Chen,CJ-Butler,S-Qiao,L-Stojchevska,I-Vaskivskyi}.
Nonetheless, the understanding of the origin and nature of the IC-CDW 
is very limited at present.

Although phonon-driven CDW inevitably causes sizable 
lattice distortion (LD) in proportion to the transition temperature,
the LD below $T_{\rm IC}$ is much smaller than that below $T_{\rm C}$ in 1T-TaS$_2$
\cite{LD1,LD2}. 
Considering the importance of electron correlation in 1T-TaS$_2$, 
it is important to investigate the electron-correlation-driven
IC-CDW mechanism, although 
the derivation of ``nonmagnetic IC-CDW order'' 
is a very difficult theoretical problem.
In fact, magnetic order is always obtained
based on the Hubbard models with on-site $U$
within mean-field theories.
Thus, one may consider the existence of
large off-site bare interactions 
(such as the off-site Coulomb interaction and 
RKKY interaction \cite{RG}) comparable to $U$.
Therefore, the study of IC-CDW is very important to uncover
{\it the real Hamiltonian for 1T-TaS$_2$},
based on which the multistage CDW transition
and the superconductivity should be studied.

Recently, in various strongly correlated metals,
the electronic nematic states are actively studied by using 
beyond-mean-field theories
\cite{Chubukov,Fernandes,Fanfarillo,Onari-SCVC,Onari-FeSe,Yamakawa-FeSe,Yamakawa-Cu,Tsuchiizu1,Tazai-RG,Tsuchiizu4,Kawaguchi,Tazai-CeB6,Tazai-BEDT,Tazai-cLC,Tazai-JPSJ,Kontani-sLC,Onari-AFN},
especially the paramagnon-interference mechanism of density-wave orders
\cite{Onari-SCVC,Onari-FeSe,Yamakawa-FeSe,Yamakawa-Cu,Tsuchiizu1,Tazai-RG,Tsuchiizu4,Kawaguchi,Tazai-CeB6,Tazai-BEDT,Tazai-cLC,Tazai-JPSJ,Kontani-sLC,Onari-AFN}.
Although the IC-CDW state in 1T-TaS$_2$ is not nematic,
it is a promising challenge to apply 
the paramagnon-interference mechanism
to this long-standing problem.

\begin{figure}[htb]
\includegraphics[width=.99\linewidth]{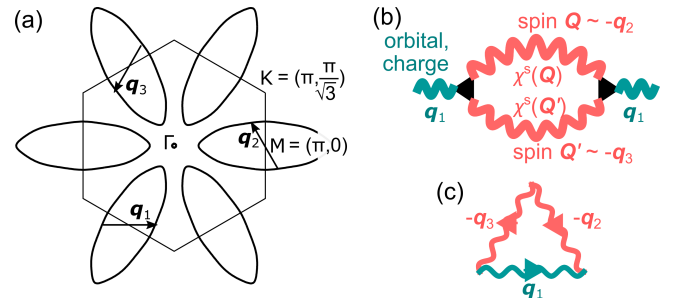}
\caption{
(a) Three electron-type FSs and the intra-FS nesting vectors $\q_n$ ($n=1,2,3$). Here, $\q_1=(0.56\pi,0)$.
(b) Paramagnon interference mechanism of charge/orbital order at wavevector $\q_1$.
(c) Momentum conservation law in the interference mechanism.
}
\label{fig:fig1}
\end{figure}

In this paper, 
we present a natural explanation for high-$T_{\rm IC}$ IC-CDW,
which is ``the parent electronic states'' of the 
exotic multistage CDW and superconductivity in 1T-TaS$_2$.
The predicted IC-CDW 
is the correlation-driven ``unconventional CDW'',
in which the CDW order parameter possesses strange
orbital-momentum-energy dependences, 
in analogy to the unconventional superconductivity.
The wavevectors of the IC-CDW state coincide with
the Fermi surface (FS) nesting vectors $\q=\q_1,\q_2,\q_3$
in Fig. \ref{fig:fig1} (a)
\cite{ARPES1,ARPES2}.
In addition, with the aid of the Ginzburg-Landau (GL) theory,
we reveal that the triple-$\q$ CDW state is stabilized.
This study provides necessary knowledge in resolving
the mysterious C-CDW state
\cite{MacMillan,Nakanishi1,Nakanishi2,Yu2017,Ikeda2019,Darancet,Chen,Ritschel}.
This theory will be useful for understanding
rich CDW states in 1T-TaS$_2$, 1T-VSe$_2$, and other TMDs.

First, we construct the first-principles
11-orbital tight-binding model of 1T-TaS$_2$, $H_0$,
composed of five $5d^1$ orbitals of Ta-ions and six $3p$ orbitals of S ions,
using the Wien2k and Wannier90 softwares.
The $d$-electron eigenfunctions in the S$_6$ octahedron 
under the trigonal distortion
are composed of one $a_{1g}$-orbital, two $e_g'$-orbitals, and two $e_g$-orbitals.
In this paper, we assign $a_{1g}$, $e_g'$, $e_g$-orbitals as
orbitals 1, (2, 3), (4, 5) in order.
The FSs are mainly composed of orbitals 1-3.
The wavefunctions of orbitals 1-5 and the model Hamiltonian 
based on experimental crystal structure \cite{crystal}
are given in the Supplemental Materials (SM) A
\cite{SM}.

We also introduce the on-site Coulomb interaction Hamiltonian, $H_U$.
It is composed of the intra (inter) orbital interaction $U$ ($U'$), 
and the exchange interaction $J=(U-U')/2$.
Below, we fix the ratio $J/U=0.1$.
(The obtained results are similar for $J/U<0.2$.)
These interaction are included into the 
spin (charge) channel interaction matrix
$\hat{\Gamma}^{s(c)}$; see SM A \cite{SM} for detail.

Here, we calculate the spin susceptibility 
using the random-phase-approximation (RPA), 
$\hat{\chi}^s(q)=\hat{\chi}^0(q)(\hat{1}-\hat{\Gamma}^s\hat{\chi}^0(q))^{-1}$,
where $q\equiv(\q,\w_l=2\pi Tl)$.
Here, $\hat{\chi}^0(q)$ is the 
irreducible susceptibility matrix
in the SM B \cite{SM}.
The spin Stoner factor $\a_S$
is the maximum eigenvalue of $\hat{\Gamma}^s\hat{\chi}^0(q)$.
The relation $\hat{\chi}^{s}(\q)\propto (1-\a_S)^{-1}$ is satisfied
when $\q$ is the nesting vector, and the magnetic order 
appears when $\a_S=1$.
The total spin susceptibility 
$\chi^{s}_{\rm tot}(\q)= \sum_{l,m=1}^5\chi^{s}_{l,l;m,m}(\q)$
exhibits broad peaks at $\q\approx\q_i$ ($i=1-3$),
as we show in Fig. S2 (a) in SM B \cite{SM}.
The components $\chi^{s}_{l,m;l,m}(\q)$ with $1\leq l,m \leq 3$
are large as shown in Fig. S2,
because the orbitals $1\sim3$ are heavily
entangled on the FSs, as shown in Fig. S1.

However, the IC-CDW without magnetization 
cannot be explained by the RPA because charge Stoner factor 
is always smaller than $\a_S$ in the RPA
\cite{Onari-SCVC}.
To explain the IC-CDW state,
we study the charge-channel susceptibility $\chi^{\rm DW}(\q)$
due to the higher-order vertex corrections (VCs),
based on the density-wave (DW) equation method
\cite{Onari-FeSe,Kawaguchi}.
(The VCs are dropped in the RPA.)
Figure \ref{fig:fig1} (b) is an Aslamazov-Larkin (AL) type VC
for $\chi^{\rm DW}(\q)$,
which is proportional to the convolution of paramagnons
$C_\q\sim \sum_\p \chi^s(\q_1+\p)\chi^s(\p)$.
We will show that the AL type VC induces the IC-CDW order at 
$\q=\q_i$ ($i=1,2,3$),
since $C_\q$ is large at $\q=\q_i$
due to the momentum conservation in Fig. \ref{fig:fig1} (c).

Here, we introduce the
linearized charge-channel DW equation 
\cite{Onari-FeSe,Kawaguchi}: 
\begin{eqnarray}
\lambda_{\q}f_\q^{L}(k)&=& -\frac{T}{N}\sum_{p,M_1,M_2}
I_\q^{L,M_1}(k,p) 
\nonumber \\
& &\times \{ G(p_-)G(p_+) \}^{M_1,M_2} f_\q^{M_2}(p) ,
\label{eqn:DWeq}
\end{eqnarray}
where $\k_\pm \equiv \k \pm \q/2$, 
$k\equiv (\k,\e_n)$ and $p\equiv (\p,\e_m)$
($\e_n$, $\e_m$ are fermion Matsubara frequencies).
$L\equiv (l,l')$ and $M_i$ represents the pair of $d$-orbital indices.
$\lambda_{\q}$ is the eigenvalue and $f_\q^L(k)$ is the 
Hermite form factor.
The former represents the 
instability of the DW fluctuations at wavevector $\q$,
which reaches unity when the long-range order is established.
The latter is the general charge-channel order parameter:
$f_\q^{l,l'}(\k)=\sum_{\s}\langle c_{\k_+,l,\s}^\dagger c_{\k_-,l',\s} \rangle
-\langle c_{\k_+,l,\s}^\dagger c_{\k_-,l',\s} \rangle_0$.
The DW equation (\ref{eqn:DWeq}) is interpreted as the 
``charge-channel electron-hole pairing equation''
with the pairing interaction $I_\q^{L,M}(k,p)$.

\begin{figure}[htb]
\includegraphics[width=.99\linewidth]{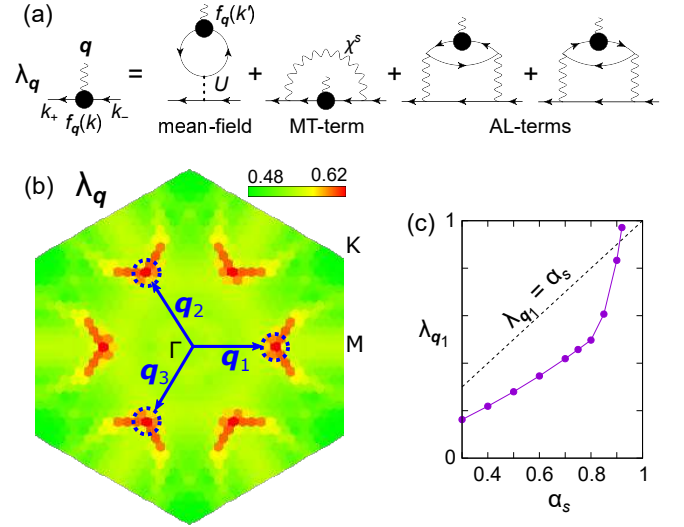}
\caption{
(a) Diagrammatic expression of the DW equation.
(b) Obtained eigenvalue of the DW equation $\lambda_\q$
for $T=40$meV and $\a_S=0.85$.
$\lambda_\q$ shows the peaks at the intra-FS nesting vectors
$q=\q_n$ ($n=1,2,3$), consistently with experiments.
Here, $\q_1=(0.56\pi,0)$.
(c) Eigenvalue $\lambda_{\q_n}$ as function of $\a_S$.
It reaches unity for $\a_S\gtrsim0.90$.
}
\label{fig:fig2}
\end{figure}

$I_\q^{L,M}$ at $\q=0$
is given by the Ward identity $-\delta\Sigma^L(k)/\delta G^M(k')$
that is composed of 
one single-magnon exchange Maki-Thompson (MT) term
and two double-magnon interference AL terms;
see Fig. \ref{fig:fig2} (a).
Here, we set $T=0.04$eV and $U=3.87$eV ($\a_S=0.85$).
(Since $d$-electron weight in the DOS at the Fermi level is about 70\%, 
$U$ is reduced to $\sim2.9$eV in the $d$-orbital Hubbard model.)
The analytic expression of the MT and AL terms
are explained in the SM C \cite{SM}.
The AL terms are proportional to the convolution of paramagnons,
$C_\q$, so they become important when $\a_S$ approaches unity
\cite{Onari-SCVC,Kawaguchi}.
Their essential role has been
revealed by the functional-renormalization-group (fRG) study
in which higher-order VCs are produced in an unbiased way
\cite{Tsuchiizu1,Tsuchiizu4}.
In contrast, the MT term is important for the 
superconducting gap equation and for the transport phenomena
\cite{Kontani-rev}.

Figure \ref{fig:fig2} (b) shows
the obtained eigenvalue of the DW equation $\lambda_\q$
for $T=0.04$eV and $\a_S=0.85$;
We see that $\lambda_\q$ has six peaks at the nesting vector 
($\q=\pm\q_n$, $n=1,2,3$).
As shown in Fig. \ref{fig:fig2} (c),
the eigenvalue $\lambda_{\q_1}$ increases with $U$,
and it exceeds $\a_S$ and reaches unity for $\a_S\gtrsim0.90$.
The susceptibility of the DW is given as
$\chi^{\rm DW}(\q)\propto 1/(1-\lambda_\q)$.
Thus, the obtained results are consistent with the 
IC-CDW order at $\q=\pm\q_n$ without magnetization in 1T-TaS$_2$.

We stress that $T_{\rm IC}$ is enlarged by phonons
in case that the form factor of the electron-phonon interaction
is equal to the DW form factor in symmetry
\cite{Kontani-softening-phonon}. 
As discussed in Ref. \cite{Kontani-softening-phonon},
the ``total DW susceptibility'' is given as
$\chi_{\rm tot}^{\rm DW}(\q) =\chi^{\rm DW}(\q)/(1-2g\chi^{\rm DW}(\q))$,
where $g (>0)$ is the phonon-induced attraction.
In fact, In Fe-based superconductors, 
the nematic transition temperature is raized by $\sim50$K 
due to the $B_{1g}$ phonon mode.
Similar phonon-assisted increment of $T_{\rm IC}$
is expected in 1T-TaS$_2$.

\begin{figure}[htb]
\includegraphics[width=.99\linewidth]{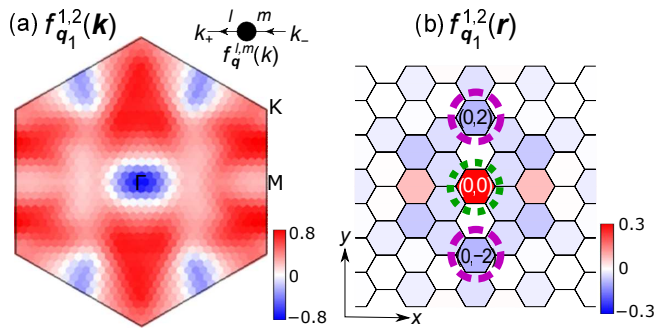}
\caption{
(a) Real-part of obtained
form factor $f_{\q_1}^{1,2}(\k)$, which
is the most significant component for the IC-CDW at $\q=\q_1$.
(b) Fourier transformation of the form factor 
$f_{\q_1}^{1,2}(\r)$.
The signal at $\r=\bm{0}$
represents the local orbital+charge order 
with respect to orbitals 1 and 2,
and the signal at $\r\ne\bm{0}$ exhibits the non-local bond order.
The strongest bond order is at $\r=(0,\pm2)$ 
marked by pink broken circles. 
}
\label{fig:fig3}
\end{figure}

Here, we discuss the nature of the form factor at $\q=\q_i$.
In Fig. \ref{fig:fig3} (a), we show the obtained real-parts of 
$f_{\q_1}^{1,2}(\k)$ at the lowest Matsubara frequency.
It is the largest intra-orbital component of the form factor
${\hat f}_{\q_1}(\k)$.
The largest intra-orbital component is $f_{\q_1}^{1,1}(\k)$,
which is exhibited in Fig. S3 (a) in the SM B \cite{SM}.
Other important form factors $f_{\q_1}^{l,m}(\k)$ 
with $1\leq l,m \leq3$
are shown in Figs. S3 (b)-(e).
The parity of the obtained order ($\k \leftrightarrow -\k$) is even.
The off-diagonal component $f_{\q_1}^{1,2}$ 
induces the orbital polarization;
$(|1\rangle,E_1) \rightarrow (|1\rangle+c|2\rangle,E_1+\Delta E)$ 
and
$(|2\rangle,E_2) \rightarrow (|2\rangle-c|1\rangle,E_2-\Delta E)$ 
with $|c|\le1$.
It also induces the finite charge order ($\delta n$)
since $E_1\ne E_2$ in the present model,
as shown Fig. S1 (b) in SM A \cite{SM}.
Interestingly,  Fig. \ref{fig:fig3} (a) has sign reversal,
which is very different from the conventional CDW 
with nearly constant form factor.
Similar ``sign reversing form factor'' is
observed in the nematic phase in FeSe by ARPES measurement
\cite{Shimojima-FeSe},
and it is satisfactorily explained as the 
paramagnon interference mechanism 
\cite{Onari-FeSe}.

Next, we examine the real space DW structure with $\q=\q_n$.
For this purpose, we perform the Fourier transformation 
of the form factor:
\begin{eqnarray}
\tilde{f}^{l,m}(\r_i,\r_j)&=& {\rm Re}\left\{ 
f_{\q_n}^{l,m}(\r_i-\r_j) e^{i\q_n\cdot(\r_i+\r_j)/2+i\theta} \right\}
\label{eqn:form-rr}
 \\
f_{\q_n}^{l,m}(\r) &=& \frac1N \sum_{\k}f_{\q_n}^{l,m}(\k) e^{i\r\cdot\k}
\label{eqn:form-r}
\end{eqnarray}
where $\r_i$ is the real space position of site $i$,
and $\theta$ is arbitrary phase factor.
Here, $\tilde{f}^{l,m}(\r_i,\r_i)$
represents the local charge and/or orbital order at $\r_i$,
and $\tilde{f}^{l,m}(\r_i,\r_j)$ with $i\ne j$
is the bond order ({\it i.e.}, the modulation of the hopping integral)
between $\r_i$ and $\r_j$.
Figure \ref{fig:fig3} (b) is the obtained $f_{\q_1}^{1,2}(\r)$:
Its large magnitude at $\r=\bm{0}$
represents the orbital order with respect to 
$|1\rangle\pm|2\rangle$.
In addition, $f_{\q_1}^{1,2}(\r)$ shows large values for $\r\ne\bm{0}$.
Thus, both orbital order and bond order coexist in the 
obtained IC-CDW state.
The trace of the form factor,
$f_{\q_1}^{\rm tr}(\r) \equiv \sum_l f_{\q_1}^{l,l}(\r)$,
is shown in Fig. S3 (f) in the SM B.
The large value of $f_{\q_1}^{\rm tr}(\r)$ at $\r=\bm{0}$
means the emergence of the local charge order.
To summarize, the IC-CDW in 1T-TaS$_2$ is identified as
the combination of the charge/orbital/bond order in the present study.

In the obtained IC-CDW state,
essentially all elements of the form factor $f_{\q_n}^{l,m}$
in the $a_{1g}+e_g'$ orbital space ($l,m=1\sim3$) are large.
This fact indicates that all $a_{1g}+e_g'$-orbital states
cooperatively magnify the eigenvalue.
In order to identify the major order parameter,
we solve the DW equation (\ref{eqn:DWeq}) for $\q=\q_1$
by considering only $f_{\q_1}^{1,2}$ and $f_{\q_1}^{2,1}$,
by setting other elements zero.
In this case, the obtained form factor $f_{\q_1}^{1,2}(\k)$
is very similar to Fig. S3 (a),
and $\lambda_{\q_1}$ is reduced just to 86\% from Fig. \ref{fig:fig2} (b).
In contrast, 
the eigenvalue becomes very small 
if only diagonal elements, $f_{\q_1}^{l,l}$, are considered.
Therefore, off-diagonal form factor $f_{\q_1}^{1,2}$ is
the main order parameter of the IC-CDW state.

\begin{figure}[htb]
\includegraphics[width=.99\linewidth]{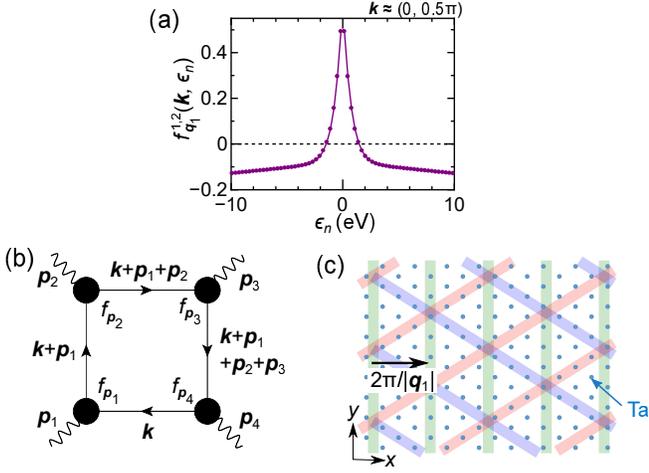}
\caption{
(a) ${\rm Re}f_{\q_1}^{1,2}(\k,\e_n)$ 
as function of $\e_n$ at $\k=(0,0.5\pi)$.
(b) Diagrammatic expression of the fourth-order term 
in the GL equation.
(c) Triple-$\q$ IC-CDW state in real-space.
}
\label{fig:fig4}
\end{figure}

In the present mechanism, 
the orbital+charge DW order due to $f_{\q_1}^{1,2}\ne0$ originates from the
interference among spin fluctuations at $\q=\q_2$ and $\q=\q_3$,
as depicted in Fig. \ref{fig:fig1} (c).
Mathematically, $f_{\q_1}^{1,2}$ is derived from the kernel $I_{\q_1}^{L,M}$.
However, local net charge order ($\delta n$) is energetically unfavorable
due to the mean-field term ($\sim U\delta n$)
in the first term of Fig. \ref{fig:fig2} (a).
In fact, in Fe-based and cuprate superconductors,
the bond and orbital orders with $\delta n=0$
appear since they are not prohibited by on-site $U$
\cite{Onari-FeSe,Tsuchiizu4}.
To understand why net charge order is obtained 
in 1T-TaS$_2$ model, we examine the energy-dependence of the form factor.
Figure \ref{fig:fig4} (a) shows the frequency ($\e_n$)
dependence of ${\rm Re}f_{\q_1}^{1,2}(\k,\e_n)$ near the 
van-Hove singular point $\k=(0,0.5\pi)$.
Similar sign reversal appears in other elements of the form factor.
This sign reversing form factor is very similar to the 
sign reversing gap function in the $s$-wave superconductors with $U\ne0$,
known as the ``retardation effect'' that drastically reduces the
depairing by $U$.
Thus, net charge order in the IC-CDW state in 1T-TaS$_2$,
which is very unusual in metals with large $U$, 
is stabilized by the retardation effect. 
To summarize, the predicted ``unconventional CDW state''
in 1T-TaS$_2$ is characterized by the orbital selective form factor
with strange sign reversals in the momentum and energy spaces.

Finally, 
we explain the ``triple-$\q$ CDW state'' in 1T-TaS$_2$,
which is the uniform coexisting state of
the three order parameters with $\q=\q_1$, $\q_2$, $\q_3$.
For this purpose, we construct a simple Ginzburg-Landau Free energy
for the CDW order 
$\vec{\Phi}(k)=(\eta_1 f_{\q_1}(k),\ \eta_2 f_{\q_2}(k), \ \eta_2 f_{\q_2}(k))$,
where $\eta_i$ is the real order parameter 
and $f_{\q_i}(k)$ is the normalized form factor.
Then, the free energy is given by
$F=a_0(|\eta_1|^2+|\eta_2|^2+|\eta_3|^2)+b_0(|\eta_1|^4+|\eta_2|^4+|\eta_3|^4)
+c_0(|\eta_1\eta_2|^2+|\eta_2\eta_3|^2+|\eta_3\eta_1|^2)$,
where $a_0\propto T-T_{\rm IC}$.
The fourth-order coefficients $b_0$ and $c_0$ can be calculated 
by using the Green functions and form factors; see Fig. \ref{fig:fig4} (b).
The derivation of $a_0$, $b_0$ and $c_0$
in addition to the third order term $F^{(3)}=d_0 \eta_1\eta_2\eta_3$
is given in the SM D \cite{SM}.
Here, the single-$\q$ state and the triple-$\q$ state
correspond to $(\eta_1,\eta_2,\eta_3)=(\eta,0,0)$
and $(\eta,\eta,\eta)$, respectively.
It is easy to show that the triple-$\q$ condition is 
$c_0/b_0<2$ if $d_0$ is negligible.
As we show in the SM D \cite{SM},
the ratio $c_0/b_0=1.1$ is obtained by using the 
form factors in the present study.
($f_{\q_1}^{lm}$ is given in Fig. \ref{fig:fig3} (a) and Figs. S3.)
Thus, the present IC-CDW state satisfies the triple-$\q$ condition.
In contrast, 
in case of conventional CDW
form factor with $f_{\q_n}^{l,m}=\delta_{l,m}$,
the obtained ratio $c_0/b_0=3.2$ does not
satisfy the triple-$\q$ condition;
see SM D \cite{SM}.
Thus, the obtained unconventional form factor
due to the AL processes 
is indispensable to explain the triple-$\q$ IC-CDW state 
in 1T-TaS$_2$, which is schematically shown 
in Fig. \ref{fig:fig4} (c).

\begin{figure}[htb]
\includegraphics[width=.99\linewidth]{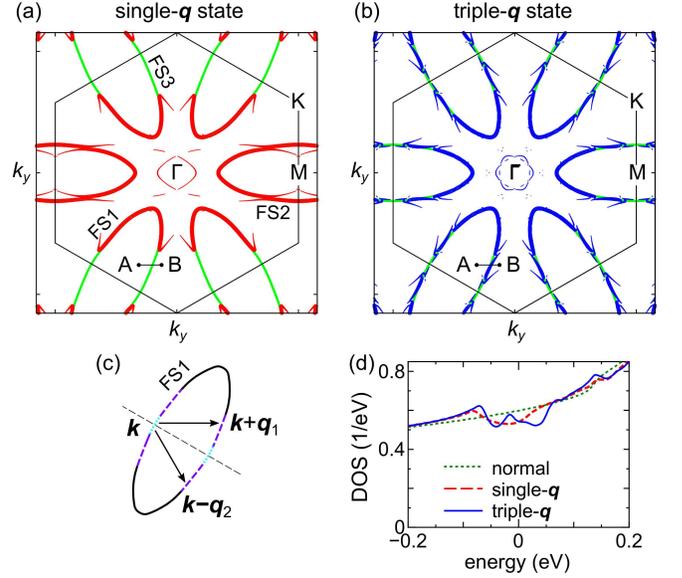}
\caption{
(c) Unfolded FS in the single-$\q_1$ state
and (d) that in the triple-$\q$ state.
(c) Reason for the FS recovery in the triple-$\q$ CDW,
which occurs when the state $|\k\rangle$ hybridizes with $|\k+\q_1\rangle$
and $|\k-\q_2\rangle$ simultaneously.
(d) DOS as function of $\e-E_{\rm F}$.
We introduced BCS-type cutoff $\w_c=4f^{\rm max}$.
}
\label{fig:fig5}
\end{figure}

Here, we calculate the electronic states below $T_{\rm IC}$
based on the $4\times4$ cluster tight-binding model
with finite DW order given in Eq. (\ref{eqn:form-rr}).
We make the wavevector of the DW order
$\q_1=(0.5\pi,0)$ by introducing 20\% hole-doping.
Although the folded FS under the CDW state is very complex,
it can be ``unfolded'' to the original BZ by restoring the 
translational symmetry of the spectral function 
\cite{Ku}.
Figures \ref{fig:fig5} (a) and (b) show the obtained unfolded FS 
under the single-$\q$ and triple-$\q$ CDW states, respectively,
by setting $f^{\rm max}\equiv \max_{l,m,\k} \{f_{\q_1}^{l,m}(\k)\}=88$meV.
In the single-$\q_1$ case, sizable Fermi arc appears
in FS1,3 due to the band-folding by $\hat{f}_{\q_1}$.
In the same way, Fermi arc appears 
in FS1,2 (FS2,3) by $\hat{f}_{\q_2}$ ($\hat{f}_{\q_3}$).
The expected charge density modulation by 
$|f^{\rm tr}_{\q_1}(\r={\bm 0})|\sim0.5$ in Fig. S3 (f) is
$\delta n\sim (f^{\rm max} |f^{\rm tr}_{\q_1}(\r={\bm0})|)N(0)\sim 0.02$.

Counter-intuitively,
the size of the Fermi arc in the triple-$\q$ case
in Fig. \ref{fig:fig5} (b) is much reduced,
where we set $f^{\rm max}=(88/\sqrt{2})$meV
because $b_0\approx c_0$ in the present study;
see SM D \cite{SM}.
To understand the reason,
we consider the hybridization between Fermi momenta 
$\k$, $\k+\q_1$ and $\k-\q_2$
in Fig. \ref{fig:fig5} (c).
In the FS reconstruction by two form factors
$\hat{f}_{\q_1}$ and $\hat{f}_{\q_2}$,
the state $|\k\rangle$ hybridizes with $|\k+\q_1\rangle$
and $|\k-\q_2\rangle$ at the same time.
Since $\hat{f}_{\q_1}\sim \hat{f}_{\q_2}$,
one eigenstate $|\k+\q_1\rangle-|\k-\q_2\rangle$ 
is unhybridized and therefore gapless.
(For general hybridization potentials, 
one of the three bands always remains unhybridized.)
For this reason, after the unfolding, the Fermi arc structure
around $\k+\q_1$ and $\k-\q_2$ in Fig. \ref{fig:fig5} (a)
is recovered, as shown in Fig. \ref{fig:fig5} (b).
Also, the spectral recovery in the unfolded bandstructure 
is explain in the SM B \cite{SM}.
This hallmark in the triple-$\q$ CDW state could be observed
by high-resolution ARPES study.
Figure \ref{fig:fig5} (d) shows the obtained density-of-states (DOS).
The pseudogap at $E_{\rm F}$ in the triple-$\q$ CDW is small
by reflecting the short Fermi arc in Fig. \ref{fig:fig5} (b).
This result is consistent with experimental 
good metallic state below $T_{\rm IC}$.

In summary,
we succeeded in explaining the triple-$\q$ IC-CDW 
in 1T-TaS$_2$ in terms of the ``unconventional CDW'',
in which the form factor has strange
orbital-momentum-energy dependences.
Thanks to the present paramagnon interference mechanism,
the triple-$\q$ IC-CDW state is naturally understood
based on a simple Hubbard model with on-site $U$,
without introducing any nonlocal interactions.
The same mechanism would be applicable for 1T-VSe$_2$ and other TMDs.
Based on the knowledge on the IC-CDW state obtained by this study,
it would be useful to develop Ginzburg-Landau theory 
to understand the NC- and C-CDW states.

\acknowledgements
We are grateful to R. Tazai for useful discussions.
This study has been supported by Grants-in-Aid for Scientific
Research from MEXT of Japan.
This work was supported by the ``Quantum Liquid Crystals'' 
No. JP19H05825 KAKENHI on Innovative Areas from JSPS of Japan, 
and JSPS KAKENHI (JP18H01175, JP17K05543, JP20K03858).



\clearpage
\newpage


\makeatletter
\renewcommand{\thefigure}{S\arabic{figure}}
\renewcommand{\theequation}{S\arabic{equation}}
\makeatother
\setcounter{figure}{0}
\setcounter{equation}{0}
\setcounter{page}{1}
\setcounter{section}{1}

\begin{widetext}
\begin{center}
{\bf \large 
[Supplementary Material] \\
\vspace{3mm}
{\large
Unconventional orbital-charge density wave mechanism
in transition metal dichalcogenide 1T-TaS$_2$
}
}%
\end{center}

\begin{center}
Toru Hirata, Youichi Yamakawa, Seiichiro Onari, and Hiroshi Kontani
\end{center}

\begin{center}
\textit{Department of Physics, Nagoya University, Nagoya 464-8602, Japan}
\end{center}

\end{widetext}

\section{A: Model Hamiltonian}

\begin{figure}[htb]
\includegraphics[width=.99\linewidth]{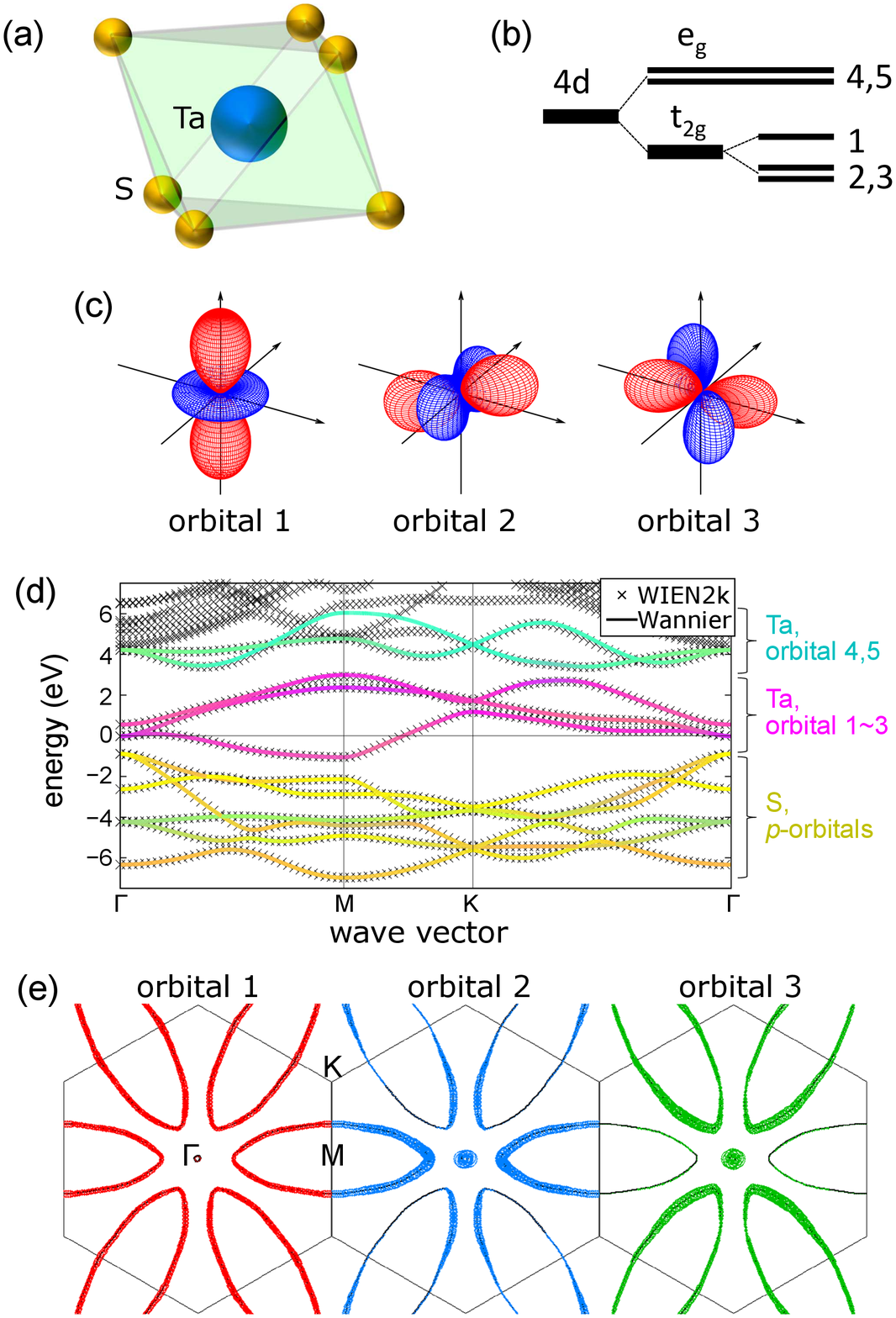}
\caption{
(a) TaS$_6$ octahedron.
(b) $5d$ orbital levels in TaS$_6$ octahedron.
(c) Real-space pictures of orbitals 1, 2, and 3.
(d) Band structure of 1T-TaS$_2$ given by the 
present multiorbital model.
The original bandstructure by WIEN2k software is also shown.
(e) Weights of the orbitals 1-3 on the FS.
}
\label{fig:figS1}
\end{figure}

Here, we construct the multiorbital Hubbard model for 1T-TaS$_2$.
In this metal compound, the TaS$_2$-plane gives
the almost perfect two-dimensional bandstructure.
Figure \ref{fig:figS1} (a) shows the TaS$_6$ octahedron
that is the building block of the TaS$_2$-plane.
The $5d$ orbital levels in TaS$_6$ octahedron are shown in 
Fig. \ref{fig:figS1} (b).
When the octahedron is regular,
the $5d$ orbitals split into 
$t_{2g}$ orbitals (1-3) and $e_g$ orbitals (4,5), respectively.
However, $t_{2g}$ orbitals split into orbital 1 and orbitals 2,3 
due to the finite trigonal distortion in real 1T-TaS$_2$.
Then, the wavefunction of orbitals 1-5 are given as
\begin{eqnarray}
{\rm orbital \ 1}&:& |d_{3z^3-r^2}\rangle
\label{eqn:S-orb1}
 \\
{\rm orbital \ 2}&:& a |d_{x^2-y^2}\rangle +b|d_{xz}\rangle
 \\
{\rm orbital \ 3}&:& a |d_{xy}\rangle -b|d_{yz}\rangle
 \\
{\rm orbital \ 4}&:& a |d_{yz}\rangle +b|d_{xy}\rangle
 \\
{\rm orbital \ 5}&:& a |d_{xz}\rangle -b|d_{x^2-y^2}\rangle
\label{eqn:S-orb5}
\end{eqnarray}
where $a^2+b^2=1$ and $a>b>0$.

Now, we calculate the bandstructure of 1T-TaS$_2$
using the Wien2k software, by referring 
the experimental crystal structure 
\cite{S-crystal}.
Next, we construct the 11 orbital tight-binding model
using the Wannier90 software, 
by taking account of the 4$d$-orbital wavefunction 
in Eqs. (\ref{eqn:S-orb1})-(\ref{eqn:S-orb5}).
The orthogonality of the $5d$ orbital is maintained
by setting $a=0.847$ ($b=0.532$).
The wavefunctions of 1-3 orbitals are shown in 
Fig. \ref{fig:figS1} (c).

Figure \ref{fig:figS1} (d) exhibits the
bandstructure of the obtained model,
which reproduces the original bandstructure by WIEN2k satisfactorily.
The conduction bands are mainly composed of 1-3 orbitals.
Figure \ref{fig:figS1} (e) shows the weight of 
each $d$-orbital on the FS.
We see that three $d$-orbitals are heavily entangled
in many parts of the FS.
Due to this fact, all the form factors $f_\q^{l,m}(\k)$ 
with $l,m=1\sim3$ become large in the DW equation analysis.
This fact is favorable for high temperature transition 
temperature $T_{\rm IC}$.

We note that the small hole-pockets around $\Gamma$ point 
in Figs. \ref{fig:figS1} (d) and (e) are not observed 
by recent ARPES study \cite{SDimer-Mott2020},
and they do not exist in the recent first principles study \cite{SYu2017}.
Therefore, in order to eliminate their artificial contributions,
we multiply the Green function $G(k)$ with
the Heaviside step function $\Theta(|\k|-k_0)$ with $k_0=0.15\pi$ 
in the present numerical study in the main text.

Next, we explain the multiorbital Coulomb interaction.
The matrix expression of the 
spin-channel Coulomb interaction is
\begin{equation}
\Gamma_{l_{1}l_{2},l_{3}l_{4}}^s = \begin{cases}
U, & l_1=l_2=l_3=l_4 \\
U' , & l_1=l_3 \neq l_2=l_4 \\
J, & l_1=l_2 \neq l_3=l_4 \\
J' , & l_1=l_4 \neq l_2=l_3
\end{cases}
\end{equation}
The matrix expression of the 
charge-channel Coulomb interaction is
\begin{equation}
\Gamma_{l_{1}l_{2},l_{3}l_{4}}^c = \begin{cases}
-U, & l_1=l_2=l_3=l_4 \\
U'-2J , & l_1=l_3 \neq l_2=l_4 \\
-2U' + J, & l_1=l_2 \neq l_3=l_4 \\
-J' , & l_1=l_4 \neq l_2=l_3
\end{cases}
\end{equation}
Here, $U$ ($U'$) is the intra-orbital (inter-orbital)
Coulomb interaction, $J$ is the Hund's coupling, 
and $J'$ is the pair hopping term.
In the main text, we assume the relations $U=U'-2J$ and $J=J'$,
and set the constraint $J/U=0.10$.
The obtained results are not sensitive to the ratio $J/U$.

\section{B: Numerical Results, Unfolded band spectrum}

The spin susceptibility in the RPA,
$\chi^s_{l,l';m,m'}(q)$, is given by
\begin{eqnarray}
\hat{\chi}^s(q)= \hat{\chi}^0(q)
(\hat{1}-\hat{\Gamma}^s\hat{\chi}^0(q))^{-1}
\end{eqnarray}
where the element of the irreducible susceptibility is
$\chi^0_{l,l';m,m'}(q)=-T\sum_k G_{l,m}(k+q)G_{m',l'}(k)$.
$G_{l,m}(k)$ is the electron Green function.

\begin{figure}[htb]
\includegraphics[width=.99\linewidth]{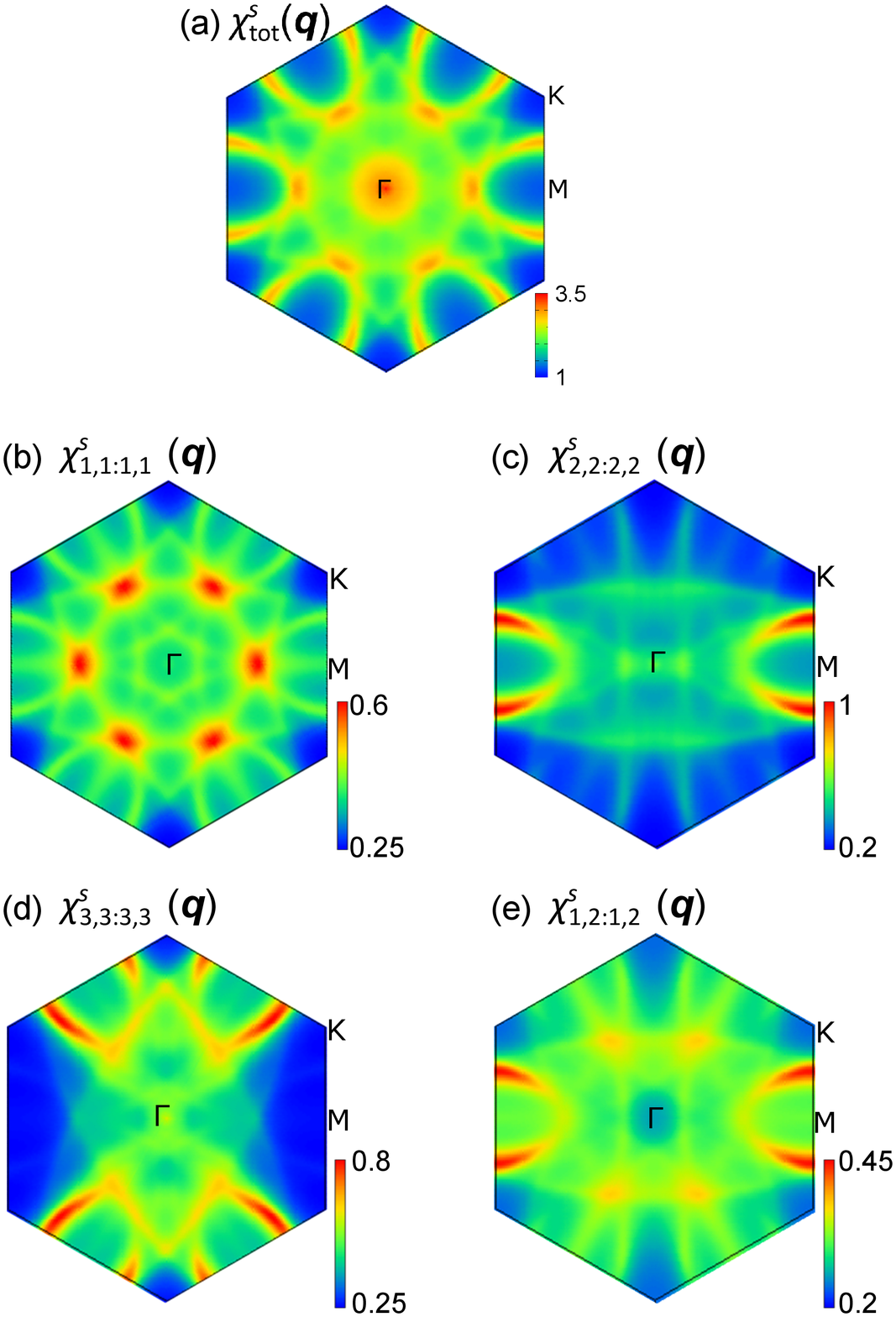}
\caption{
Obtained spin susceptibilities for $\a_S=0.85$:
(a) $\chi^s_{\rm tot}(\q)$,
(b) $\chi^s_{1,1;1,1}(\q)$,
(c) $\chi^s_{2,2;2,2}(\q)$,
(d) $\chi^s_{3,3;3,3}(\q)$, and 
(e) $\chi^s_{1,2;1,2}(\q)$.
}
\label{fig:figS2}
\end{figure}

In the present model, 
$\chi^s_{l,l';m,m'}(q)$ becomes large for
$l,l',m,m'=1\sim3$.
Figure \ref{fig:figS2} (a) shows the 
total spin susceptibility 
$\chi^s_{\rm tot}(\q)\equiv \sum_{l,m}\chi^s_{l,l;m,m}(\q)$.
We also display some important elements of $\chi^s_{l,l';m,m'}(q)$
in Figs. \ref{fig:figS2} (b)-(e)

\begin{figure}[htb]
\includegraphics[width=.99\linewidth]{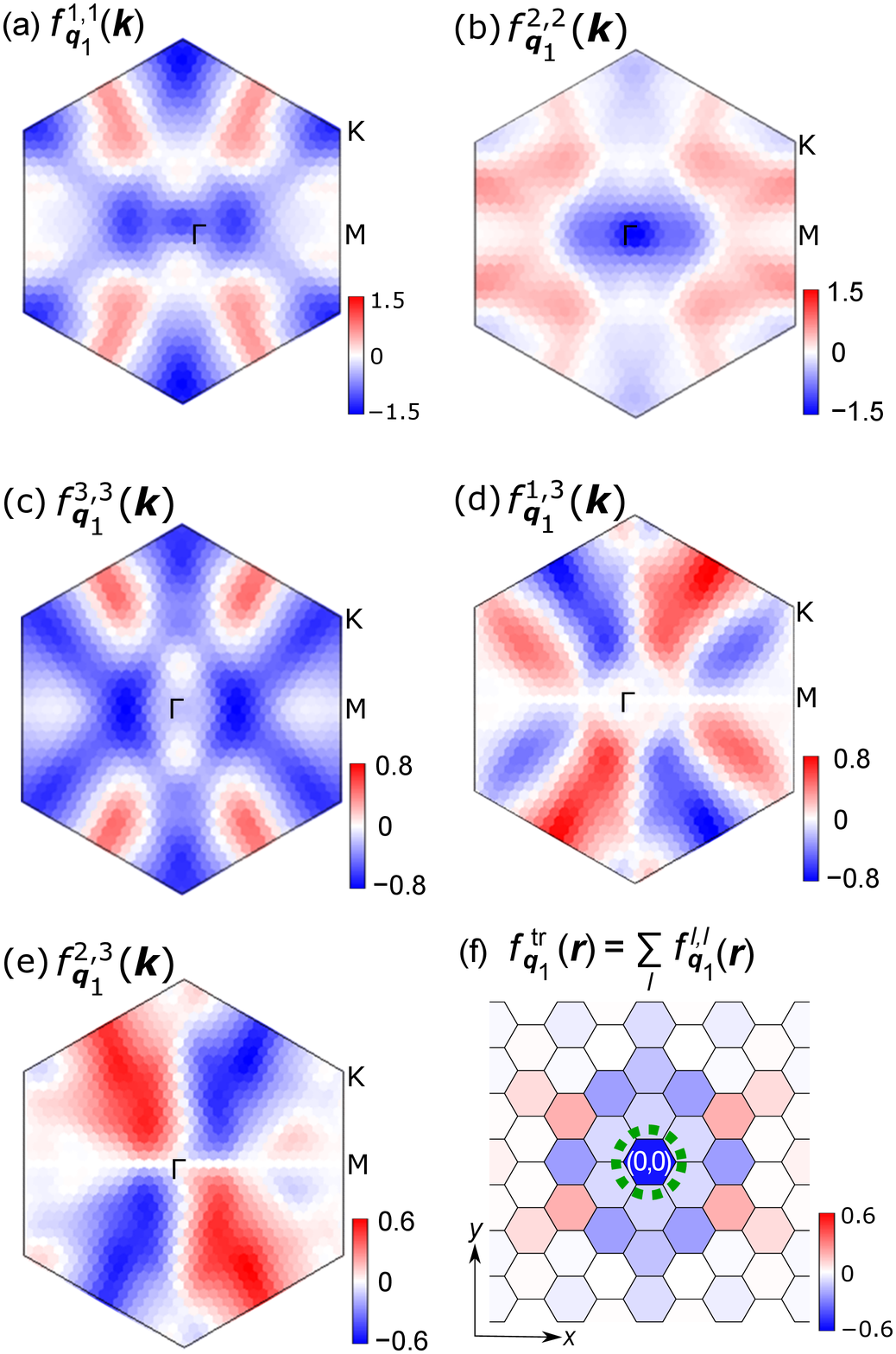}
\caption{
Real-parts of the obtained form factors $f^{l,m}_{\q_1}(\k)$
in the case of $\a_S=0.85$:
(a) $f^{1,1}_{\q_1}(\k)$,
(b) $f^{2,2}_{\q_1}(\k)$,
(c) $f^{3,3}_{\q_1}(\k)$,
(d) $f^{1,3}_{\q_1}(\k)$, 
(e) $f^{2,3}_{\q_1}(\k)$.
(f) $f_{\q_1}^{\rm tr}(\r)\equiv \sum_l f_{\q_1}^{l,l}(\r)$.
}
\label{fig:figS3}
\end{figure}

Next, we explain the form factor obtained by the DW equation.
The major form factor in the present study is $f_{{\bm q}_1}^{1,2}$,
which is shown in Fig. 3 (b) in the main text.
It mainly originates from the AL process due to large value of 
$\chi^s_{1,1;1,1}(\q)$ at $\q\sim\q_1$
and that of $\chi^s_{1,2;1,2}(\q)$ at $\q\sim\bm{0}$.
In Figs. \ref{fig:figS3} (a)-(e),
we show other form factors $f_{{\bm q}_1}^{l,m}$ for $l,m=1\sim3$.
Thus, both diagonal and off-diagonal form factors 
with respect to orbitals $1\sim3$ take large values
in the present numerical study.
In Fig. \ref{fig:figS3} (f),
the signal at $\r=\bm{0}$
represents the presence of the local charge modulation.

Thus, all elements of $f_{{\bm q}_1}^{l,m}$ with $l,m=1\sim3$ are 
large and important.
The main form factor $f_{{\bm q}_1}^{1,2}$ induces not only the 
orbital polarization, but also net charge modulation,
because orbitals 1 and 2 are not degenerated.
For this reason, net charge modulation in Fig. \ref{fig:figS3} (f)
exists in the present unconventional CDW state.

\begin{figure}[htb]
\includegraphics[width=.99\linewidth]{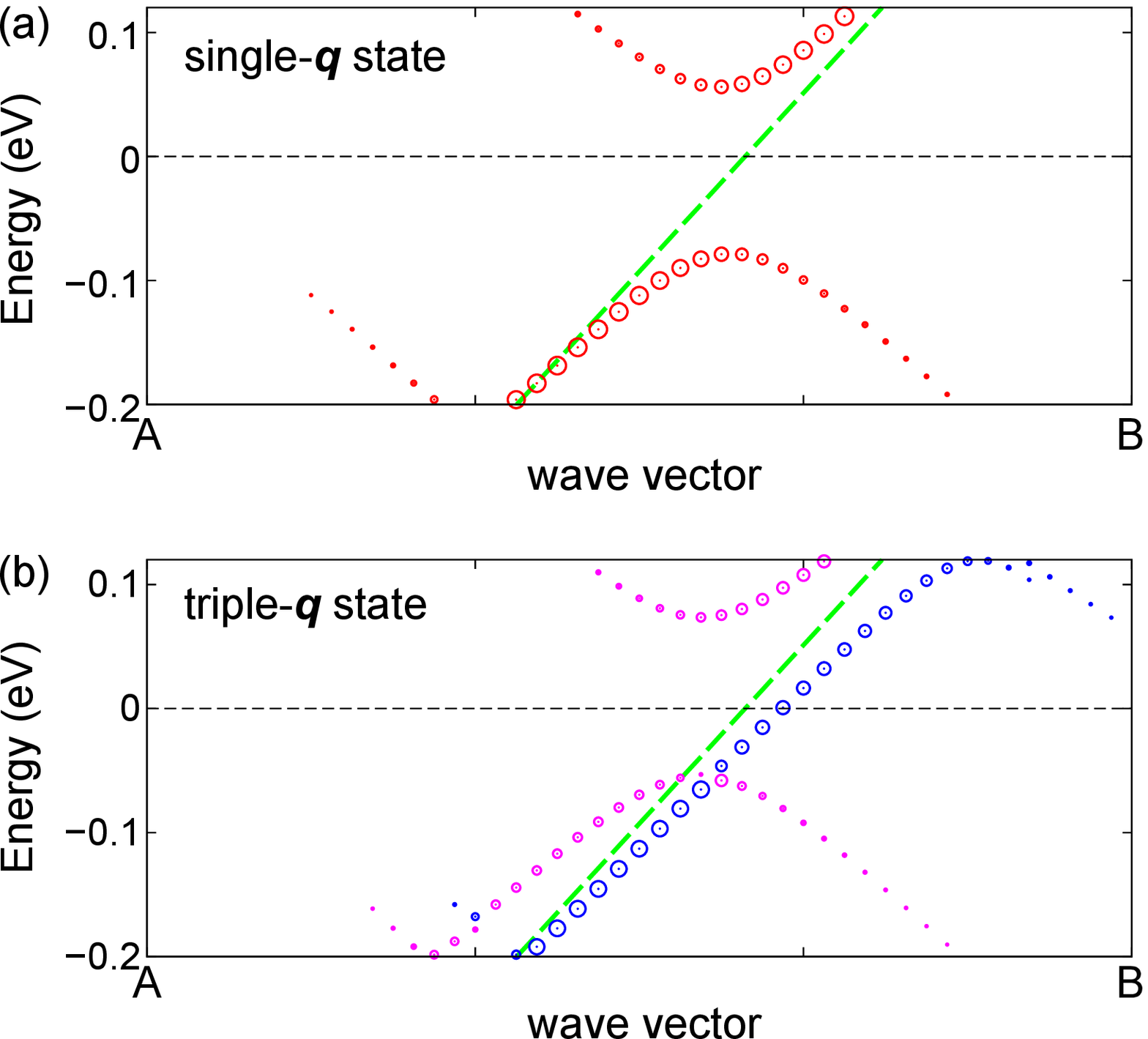}
\caption{
Unfolded bandstructures in the (a) single-$\q$ CDW state
and (b) triple-$\q$ CDW state, respectively,
along the cut A-B in Figs. \ref{fig:fig5} (a) and (b).
The green lines gives the original spectra.
The recovered spectrum in the triple-$\q$ CDW is 
stressed as blue dots in (b).
}
\label{fig:figS4}
\end{figure}

In the triple-$\q$ CDW state, both the FS and the bandstructure
are folded intricately into the folded BZ.
They can be unfolded into the original size BZ,
which correspond to the experimental results by ARPES measurements.
In the main text, we show the unfolded FSs
in both the single-$\q$ and triple-$\q$ CDW states.
In Figs. \ref{fig:figS4} (a) and (b),
we show the unfolded bandstructure in the (a) single-$\q$ CDW state
and (b) triple-$\q$ CDW state, respectively,
along the cut A-B in Figs. \ref{fig:fig5} (a) and (b).
(The green lines gives the original spectra.)
In the single-$\q$ CDW state,
the spectrum weight around the Fermi level disappears 
due to the band hybridization.
This missing spectrum is partially recovered in the triple-$\q$ CDW state,
as we explain in the main text.
The recovery spectrum is stressed as blue dots in Fig. \ref{fig:figS4} (b).
This hallmark in the triple-$\q$ CDW state may be observed
by high resolution ARPES study.

\section{C: Kernel function in the DW equation}

Here, we explain the kernel function in the
DW equation, $I^{l, l', m, m'}_{\q}$,
studied in the main text.
It is given by the Ward identity from the one-loop
fluctuation-exchange self-energy.
In multiorbital models, it is given as
\begin{eqnarray}
&& I^{l, l', m, m'}_{\q} (k,k')
= \sum_{b = s, c} \frac{a^b}{2} 
\Bigl[ V^{b}_{l, m; l', m'} (k-k') 
\nonumber \\
&&
-T \sum_{p} \sum_{l_1, l_2, m_1, m_2}
	V^{b}_{l, l_1; m, m_1} \left( p_+ \right) 
	V^{b}_{m', m_2; l', l_2} \left( p_- \right)
 \nonumber \\
&& \qquad\qquad\qquad \quad
\times G_{l_1, l_2} (k-p) G_{m_2, m_1} (k'-p)
 \nonumber \\
&& 
-T \sum_{p} \sum_{l_1, l_2, m_1, m_2}
	V^{b}_{l, l_1; m_2, m'} \left( p_+ \right) 
	V^{b}_{m_1, m; l', l_2} \left( p_- \right)
 \nonumber \\
&& \qquad\qquad\qquad \quad
\times G_{l_1, l_2} (k-p) G_{m_2, m_1} (k'+p) \Bigr],
\label{eqs:kernel} 
\end{eqnarray}
where $a^{s(c)} = 3$($1$), $\p_\pm \equiv \p + \q/2$, and $p = (\p,\w_l)$. 
$\hat{V}^{b}$ is the $b$-channel interaction given by 
$\hat{V}^{b} = \hat{\Gamma}^{b} + \hat{\Gamma}^{b} \hat{\chi}^{b} \hat{\Gamma}^{b}$. 
$\hat{\Gamma}^{b}$ is the matrix expression of the 
bare multiorbital Coulomb interaction for channel $b$.

The first term of Eq. (\ref{eqs:kernel}) 
corresponds to the Maki-Thompson term, 
and the second and third terms give Aslamazov-Larkin terms.

\section{D: Ginzburg-Landau equation}

Here, we construct a simple Ginzburg-Landau Free energy for the CDW order 
$\vec{\Phi}(k)=(\eta_1 f_{\q_1}(k),\ \eta_2 f_{\q_2}(k), \ \eta_3 f_{\q_3}(k))$,
where $\eta_i$ is the real order parameter 
and $f_{\q_i}(k)$ is the normalized form factor.
Then, the free energy is given by
$F=a_0(|\eta_1|^2+|\eta_2|^2+|\eta_3|^2)+b_0(|\eta_1|^4+|\eta_2|^4+|\eta_3|^4)
+c_0(|\eta_1\eta_2|^2+|\eta_2\eta_3|^2+|\eta_3\eta_1|^2)
+d_0 \eta_1\eta_2\eta_3
$
where $a_0\propto T-T_{\rm IC}$.
In the single-$\q$ state with $(\eta_1,\eta_2,\eta_3)=(\eta,0,0)$,
the order parameter below $T_{\rm IC}$ is
$\bar{\eta}=\sqrt{-a_0/2b_0}$ and $\bar{F}= -a_0^2/4b_0$.
Also, in the triplet-$\q$ state with $(\eta_1,\eta_2,\eta_3)=(\eta,\eta,\eta)$,
we obtain
$\bar{\eta}=\sqrt{-a_0/2(b_0+c_0)}$ and $\bar{F}= -3a_0^2/4(b_0+c_0)$.
Thus, the realization condition for the 
triple-$\q$ state is $c_0/b_0<2$.

From now on, we derive the 
fourth-order coefficients $b_0$ and $c_0$ 
by using the Green functions and form factors.
Here, we introduce six nesting vectors $\q_n$ ($n=1\sim6$) 
with the relation $\q_{n+3}=-\q_n$.
We also introduce the scalar order parameter $\eta_n$ 
for the form factor $f_{\q_n}$ with the relation 
$\eta_{n+3}=\eta_n^*$.
Note that the relation $f_{\q_n}=f_{\q_{n+3}}^*$ holds.
Then, the fourth term of the GL free energy is
\begin{eqnarray}
F^{(4)}
&=& \sum_{n_1,n_2,n_4,n_4} 
\eta_{n_1}\eta_{n_2}\eta_{n_3}\eta_{n_4} \cdot f^{(4)}_{n_1,n_2,n_3,n_4},
\\
f^{(4)}_{n_1,n_2,n_3,n_4} &=& c_{n_i}\cdot 
\frac{T}{4}\sum_{k,\s} 
f_{\p_1}(k-\p_1/2)f_{\p_2}(k_1-\p_2/2)
\nonumber \\
& &\times f_{\p_3}(k_2-\p_3/2)f_{\p_4}(k_3-\p_4/2)
\nonumber \\
& &\times G(k)G(k_1)G(k_2)G(k_3)
\nonumber \\
& &\times \delta_{\p_{1}+\p_{2}+\p_{3}+\p_{4},\bm{0}}
\end{eqnarray}
where $\p_i \equiv \q_{n_i}$,
$k_1=k+\p_1$, $k_2=k+\p_1+\p_2$, 
and $k_3=k+\p_1+\p_2+\p_3$.
The orbital indices in $G$ and $f$ are neglected to simplify the expression.
The diagrammatic expression is given in Fig. \ref{fig:fig4} (c).

Due to the momentum conservation law,
only the following terms remain finite:
\begin{eqnarray*}
&&\mbox{case 1. } (n_1,n_2,n_3,n_4)=(1,1,4,4), \ c_{n_i}=4,\\
&&\mbox{case 2. } (n_1,n_2,n_3,n_4)=(1,4,1,4), \ c_{n_i}=2,\\
&&\mbox{case 3. } (n_1,n_2,n_3,n_4)=(1,2,4,5), \ c_{n_i}=4,\\ 
&&\mbox{case 4. } (n_1,n_2,n_3,n_4)=(1,2,5,4), \ c_{n_i}=4,\\
&&\mbox{case 5. } (n_1,n_2,n_3,n_4)=(1,4,2,5), \ c_{n_i}=4,\\
&&\mbox{case 6. } (n_1,n_2,n_3,n_4)=(1,4,5,2), \ c_{n_i}=4,\\
&&\mbox{case 7. } (n_1,n_2,n_3,n_4)=(1,5,2,4), \ c_{n_i}=4,\\
&&\mbox{case 8. } (n_1,n_2,n_3,n_4)=(1,5,4,2), \ c_{n_i}=4.
\end{eqnarray*}
The coefficient $b_0$ ($c_0$) is given by 
the cases 1 and 2 (cases 3-8):
$b_0 = F^{(4)}_{1,4,1,4}+ F^{(4)}_{1,1,4,4}$ and
$c_0 = F^{(4)}_{1,2,4,5}+ F^{(4)}_{1,2,5,4}+F^{(4)}_{1,4,2,5}+ F^{(4)}_{1,4,5,2}+F^{(4)}_{1,5,2,4}+ F^{(4)}_{1,5,4,2}$.

Here, we perform the numerical study for $b_0$ and $c_0$.
First, we calculate them by using the 
form factors obtained in the present study, 
some of which are shown in Fig. \ref{fig:fig3} (a) 
and Fig. \ref{fig:figS3}.
At $T=0.05$eV, the obtained ratio $c_0/b_0$ is $1.1$,
and it is insensitive to the temperature.
Thus, the present IC-CDW state satisfies the triple-$\q$ state condition.
In contrast, 
in the case of the conventional CDW
form factor with $f_{\q_n}^{l,m}=\delta_{l,m}$ for any $\q_n$,
we obtain the ratio $c_0/b_0=3.2$,
which does not satisfy the triple-$\q$ state condition.
Thus, the correct form factor derived by the DW equation 
is important to explain the triple-$\q$ IC-CDW state in 1T-TaS$_2$.

Finally, we examine the smallness of the third order term 
of the GL free energy,
which exists due to the relation $\q_1+\q_2+\q_3=\bm{0}$
\cite{SMacMillan}.
It is given as
\begin{eqnarray}
F^{(3)}
&=& 
\eta_{1}\eta_{2}\eta_{3} \cdot d_0,
\\
d_0 &=& -\frac{2T}{3}\sum_{k,\s} 
f_{\q_1}(k-\q_1)f_{\q_2}(k+\q_1-\q_2/2)
\nonumber \\
& &\times
f_{\q_3}(k+\q_1+\q_2-\q_3/2)
\nonumber \\
& &\times G(k)G(k+\q_1)G(k+\q_1+\q_2)
\end{eqnarray}
which is expressed as a triangle diagram.
The orbital indices in $G$ and $f$ are neglected 
to simplify the expression.
In the numerical study, we obtain the relation
$d_0/(b_0+c_0) \approx -0.05T\ [{\rm eV}]$ 
when the form factor is normalized as 
$\max_{\k,l,m}|f_{\q_i}^{l,m}(\k,\pi T)|=1$.
Thus, the contribution from the third order term is small
at $T\sim0.04$eV and $\eta_i\sim0.1$eV.
Although the change in the triple-$\q$ condition 
due to the third order term is small,
the free energy for the triple-$\q$ state
$\eta=\eta_i$ ($i=1-3$), $F_{\rm triple}(\eta)$,
depends on the sign of $\eta$.
In the present study, the obtained $\eta$ is positive since $d_0<0$.
In Figs. 5 (b) and (d) in the main text,
we show the DOS and FS under the triple-$\q$ state
with positive $\eta$.


\end{document}